\documentclass[preprint,aps,prb,a4paper,superscriptaddress]{revtex4}

\usepackage{graphicx}
\usepackage{amsmath}
\begin{document}
\title{Bound hole states associated to individual vanadium atoms\\ 
incorporated into monolayer WSe$_2$ }

\author{Pierre~Mallet}
\email[electronic address: ]{pierre.mallet@neel.cnrs.fr}
\affiliation{Universit\'e Grenoble Alpes, Institut Neel, F-38042 Grenoble, France\\}
\affiliation{CNRS, Institut Neel, F-38042 Grenoble, France\\}

\author{Florian~Chiapello}
\affiliation{Universit\'e Grenoble Alpes, Institut Neel, F-38042 Grenoble, France\\}
\affiliation{CNRS, Institut Neel, F-38042 Grenoble, France\\}

\author{Hanako~Okuno}
\affiliation{Universit\'e Grenoble Alpes, CEA, IRIG-MEM, 38000 Grenoble, France\\}

\author{Herv\'e~Boukari}
\affiliation{Universit\'e Grenoble Alpes, Institut Neel, F-38042 Grenoble, France\\}
\affiliation{CNRS, Institut Neel, F-38042 Grenoble, France\\}

\author{Matthieu~Jamet}
\affiliation{Universit\'e Grenoble Alpes, CEA, CNRS, Grenoble INP, IRIG-SPINTEC, 38000 Grenoble, France\\}

\author{Jean-Yves~Veuillen}
\affiliation{Universit\'e Grenoble Alpes, Institut Neel, F-38042 Grenoble, France\\}
\affiliation{CNRS, Institut Neel, F-38042 Grenoble, France\\}

\date{\today}

\begin{abstract}

Doping a two-dimensional semiconductor with magnetic atoms is a possible route to induce magnetism in the material. We report on the atomic structure and electronic properties of monolayer WSe$_2$ intentionally doped with vanadium atoms by means of scanning transmission electron microscopy and scanning tunneling microscopy and spectroscopy. Most of the V atoms incorporate at W sites. These V$_W$ dopants are negatively charged, which induces a localized bound state located 140 meV above the valence band maximum. The overlap of the electronic potential of two charged V$_W$ dopants generates additional in-gap states. Eventually, the negative charge may suppress the magnetic moment on the V$_W$ dopants.

\end{abstract}

\maketitle

Magnetism in ultra-thin films is a recent development in the field of van der Waals (vdW) layered materials. For instance, the persistence of ferromagnetism in exfoliated atomic layers has been reported only in the last few years \cite{Gong Nat17, Huang Nat17}. These materials present an appealing class of systems for testing the classical models for 2D magnetism, as well as for searching for exotic electronic phases and developing  ultimately-thin devices \cite{Burch Nat18}. Being vdW layers, they can be incorporated in stacks with other metallic, semiconducting or insulating layers \cite{Ref 4} to create original devices, such as heterostructures showing giant tunneling magnetoresistance \cite{CrI3 Sci18}. Moreover, through the exchange coupling at the interface \cite{Zhao NatNano17}, such vdW heterostructures can lead to a lifting of the valley degeneracy \cite{Zhong SciAdv17} in semiconducting transition metal dichalcogenides (SC-TMDs), which is one way to get a permanent valley polarization for ÒvalleytronicsÓ applications \cite{Vitale Small18}.

Apart from exfoliating bulk materials, a promising way to prepare magnetic vdW materials in the monolayer limit is to use direct growth techniques. Such methods are extensively used to synthetise SC-TMD compounds,  with formula 1H-MX$_2$ (M=Mo or W, X=S or Se) \cite{Choi MatTod17}. Recent reports suggest  that MnSe$_2$ \cite{O'Hara NanoLett18} and VSe$_2$ \cite{Bonilla NatNano18} in the monolayer range exhibit ferromagnetism up to room temperature. However, for the former material the structure seems to change with thickness \cite{O'Hara NanoLett18}, whereas for the later the competition between different possible ground states may hinder the development of the ferromagnetic order \cite{Feng NanoLett18}. An alternative approach would be to perform magnetic doping of SC-TMD by substituting 3d transition metal atoms on the M site, to create a 2D diluted magnetic semiconductors\cite{Cheng PRB13, Mishra PRB13, Ramasubramaniam  PRB13, Fan NanoResLett16, Xie SupMic16, Wu PhysLettA18, Andriotis PRB14, ZhangRobinson NanoLett15}. 
Among the possible candidates, vanadium  is especially appealing. Its incorporation in the M sites has been demonstrated \cite{Robertson ACSNano16, Yun Condmat18}, and a ferromagnetic behavior in moderately doped samples has been reported \cite{Yun Condmat18, Duong APL19}. Moreover, ab-initio calculations predict that in the ferromagnetic state there is a significant ($\sim$ 100 meV) lifting of the valley degeneracy at the K/-K points in the Brillouin zone \cite{SinghUdo AdvMat17}, suitable for valleytronics.

In this letter, we report a study by local probe techniques of V-doped WSe$_2$ monolayers grown by molecular beam epitaxy  (MBE) on epitaxial graphene on SiC. Scanning tunneling microscopy and spectroscopy (STM/STS) is an adequate technique for detecting point defects in 2D SC-TMD and for probing their local electronic structure \cite{Schuler PRL19, Schuler ACSNano19, Barja NatCom19, LinFeenstra ACSNano18, ZhangWangLi PRL17, LeQuang 2DMater18}. With the support of additional scanning transmission electron microscopy (STEM) analysis, we identify isolated V atoms substituted on W sites (or V$_W$ dopants) in the V-doped sample, in agreement with the literature \cite{Robertson ACSNano16, Yun Condmat18}.  Our STM/STS data reveal that the substitutional V$_W$ dopants are negatively charged, and that the resulting localized repulsive potential induces a bound state of diameter $\sim$2.2 nm, located $\sim$ 140 meV above the VBM. For closely spaced V$_W$ atoms, the number and the binding energies of the bound states increase, as a consequence of the overlap of the repulsive potentials from each ionized dopant. 
The charging of the V$_W$ dopants results from two effects. Firstly, the V atoms should have an acceptor character \cite{Yun Condmat18} (as for Nb  \cite{Mukherjee PhysRevAppl17}), with an empty state just above the VBM in the neutral state as confirmed by density functional theory  \cite{Fan NanoResLett16, Wu PhysLettA18, Robertson ACSNano16, SinghUdo AdvMat17, Miao ApplSurfSci18, Mekonnen IJModPhysB18}. Secondly the electrical contact with the graphene substrate sets the position of the Fermi level of ÒpureÓ WSe$_2$ well above the valence band edge \cite{ZhangUgeda NanoLett16, ZhangFeenstra NanoLett15, LeQuang 2DMater17}, favoring a charge transfer to such acceptor state. 
Our result thus highlights the role of the environment in controlling the charge state of the V$_W$ dopant. This is a crucial point for magnetic doping of SC-TMD since the charge state of the 3d dopant atom significantly alters the ferromagnetic state of the system, as predicted by ab-initio calculations \cite{Mishra  PRB13, Robertson ACSNano16, Lin JAP14}. 


2D-WSe$_2$ flakes including a nominal amount of 3$\%$ of vanadium atoms were grown by MBE on epitaxial graphene (EG) on SiC-Si face. All the STM/STS experiments were performed at 8.5 K \cite{Ref SM}. From large scale STM images, we find that the monolayer TMD flakes present a large amount of individual species. They are abundant on our sample, but are mostly absent in the pristine TMD material \cite{Ref SM}. As shown below, we ascribe them to the successful incorporation of substitutional V$_W$ dopants in the 2D-WSe$_2$ matrix.

\begin{figure}[!h]
\begin{center}
\includegraphics[width=8.2 cm,clip]{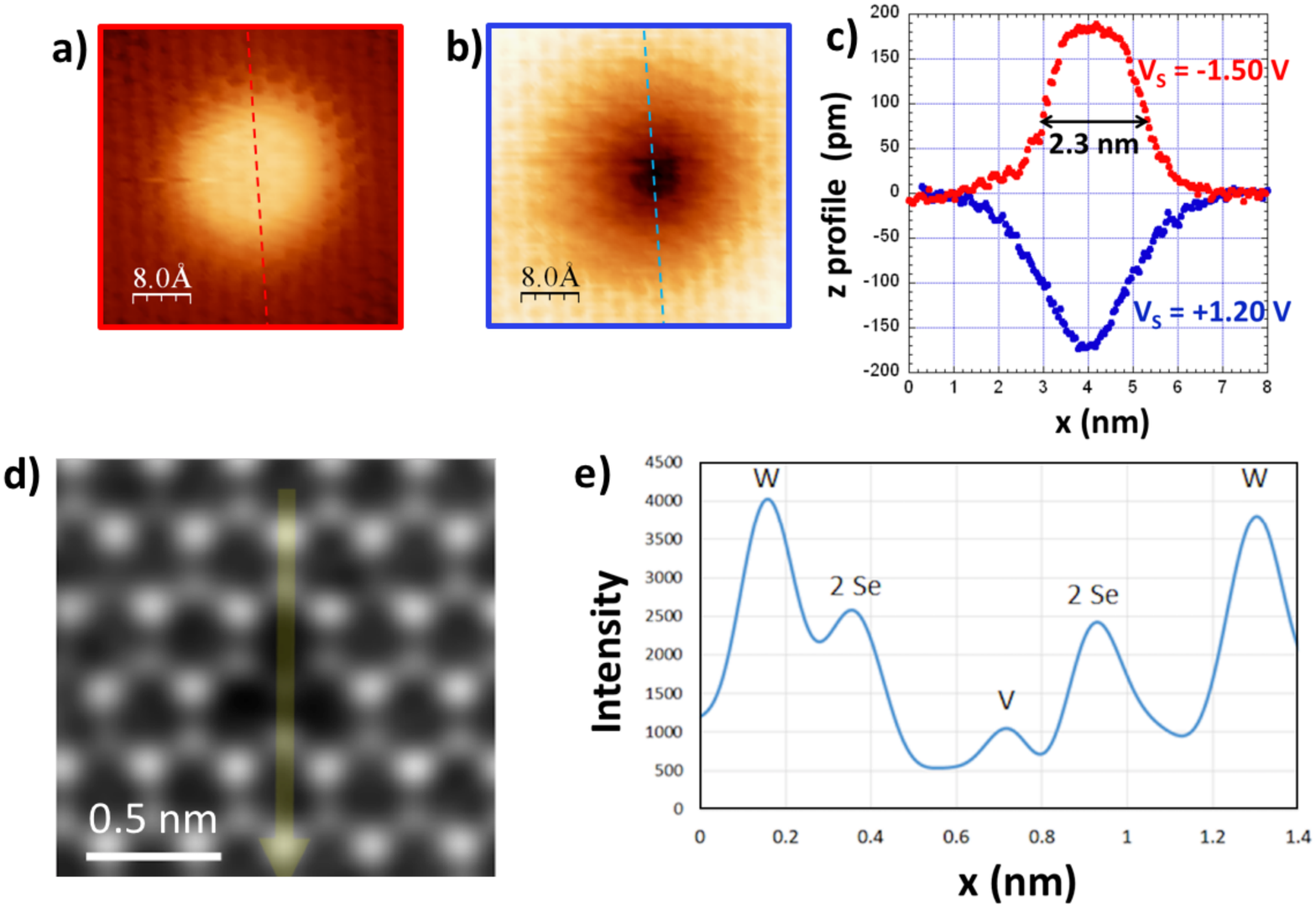}
\caption{\small STM and ADF-STEM images of individual substitutional  V$_W$ dopants incorporated in monolayer WSe$_2$. (a,b) STM images of  a single V$_W$ dopant, at sample bias $V_S$ = -1.5V (b) and $V_S$ = +1.2V (c). Image sizes: 4 $\times$ 4 nm$^2$. (c) Apparent height of the individual dopant, measured along the dashed lines drawn on (a) and (b) (red and blue curves respectively). (d) ADF-STEM image of a V-doped WSe$_2$ monolayer, centered on a substitutional V$_W$ dopant.  (e) Intensity profile along the vertical arrow in (d).}
\end{center}
\end{figure}
 
We first focus on the apparent STM diameter and height of one of these individual V dopants. On the  images at negative and positive sample bias (Fig. 1a and Fig. 1b respectively), the atomic lattice of the TMD layer is clearly resolved, with a lattice constant  0.33 nm. We find that the V dopant has a perfectly isotropic shape, with a sizable apparent diameter (2.30 $ \pm$ 0.10 nm), i.e. $\sim$ 7 times larger than the TMD lattice constant. Moreover, the  STM contrast of the dopant  is reversed when the sign of the sample bias is changed, as shown in Figs. 1a and 1b (see also Ref. \cite{Ref SM}), and on the $z$ profiles of Fig. 1c. Eventually, since the atomic pattern of the TMD shows up even in the central disk associated to the dopant, we conclude that  the surface Se atomic remains intact over the V atom. This points to an embedded dopant within the TMD layer. 

We have cross-checked  the presence of V$_W$ dopants in such MBE-grown WSe$_2$ samples, by performing angular dark field (ADF) STEM measurements on a similarly V-doped sample \cite{Ref SM}. Figure 1d is a $Z$-contrast ($Z$ is the atomic number)  image of that sample, centered on an apparently missing W. The intensity line profile (Fig. 1e) confirms the replacement of a W atom by a light atom. According to the relative $Z$-contrast between the W sites and the central site, the substitutionally positioned light atom can be considered as a vanadium single dopant \cite{Ref SM}. Similar STEM assessment has been proposed for the same system  in Ref.  \cite{Yun Condmat18}, and STEM was also previously used to evidence V atoms substituting Mo atoms in 1L-MoS$_2$ \cite{Robertson ACSNano16}. 

In order to address the full electronic structure of the individual V$_W$ dopants, we performed local tunneling spectroscopy measurements. Figure 2a presents typical $dI/dV(V)$ spectra obtained for two V$_W$ dopants (labeled B and C in the inset). These dopants belong to a monolayer WSe$_2$ flake lying on single-layer graphene (SLG).  The additional black curve in Fig. 2a is a reference spectrum recorded on the 1L-WSe$_2$ flake away from the dopants \cite{Ref SM}. As shown in Fig. 2b, this reference spectrum, when displayed on a magnified vertical scale, defines the edges of the TMD bandgap (corresponding to the bias region with vanishing tunneling conductance). We measure a bandgap of 1.90 $\pm$ 0.06 eV, which is close to the value found for monolayer WSe$_2$ on graphitized substrate  \cite{ZhangUgeda NanoLett16, ZhangFeenstra NanoLett15}. 

Examining the $dI/dV(V)$ spectra for the two V$_W$ dopants (red and purple curves in Fig. 2), we find that they are quantitatively very similar, confirming that the dopants have the same nature. They both exhibit several resonances within the valence band states of the TMD. One pronounced $dI/dV$ peak, centered at at -1.16 $\pm$ 0.02 V, shows up just below the VBM. Remarkably, a second conductance peak of weaker amplitude is also found (at -1.01 $\pm$  0.02 V as indicated by a * sign), clearly located within the TMD bandgap. We conclude that both V$_W$ individual dopants of Fig. 2 present an in-gap state ($\textit{i.e.}$ a bound state), lying 0.14 $\pm$  0.02 eV above the VBM of the 2D-TMD.

 \begin{figure}[!h]
\begin{center}
\includegraphics[width=6 cm,clip]{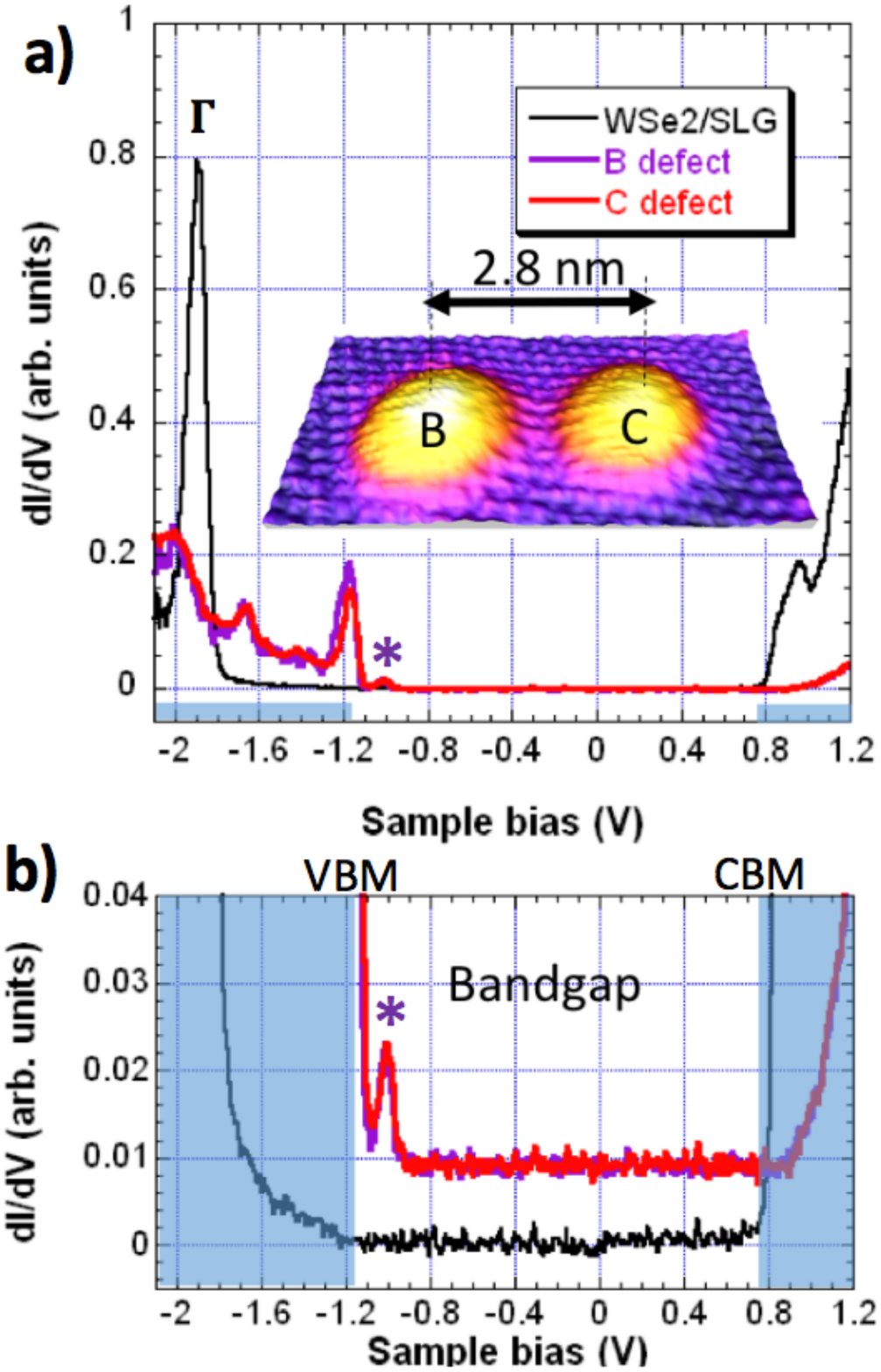}
\caption{\small STS performed at 8.5K on two neighboring V$_W$ dopants on 1L-WSe$_2$ on single-layer graphene (SLG). (a) Local $dI/dV(V)$ spectra of the two dopants labelled B and C (respectively purple and red curves). The black curve corresponds to a spectrum performed on a nearby defect-free WSe$_2$ area, the letter  $\Gamma$ indicating the energy position of the valence band maximum at  $\Gamma$  point. The inset shows a constant current STM image of the two dopants. (b) Same as (a) with a magnified vertical scale. The white box  corresponds to the TMD bandgap.  Spectra measured above the two V$_W$ dopants are almost identical, both  show a bound state (labelled *) located $\sim$0.14 eV above the VBM of the TMD.  The red and purple curves are shifted vertically for clarity. }
\end{center}
\end{figure}
 
This is a robust and systematic result, associated to most of the individual V$_W$ dopants \cite{Ref SM}. We present in Fig. 3 a full characterization of another V$_W$ dopant on a different TMD flake, here lying on a bilayer graphene region. The $dI/dV(V)$ spectrum associated to this dopant also indicates a bound state lying $\pm$ 0.14  eV above the VBM (see Ref. \cite{Ref SM}). To determine its lateral extension, we performed spectroscopic measurements along a 10 nm-long line crossing the single $V_W$ dopant (indicated by the arrow in the inset). The corresponding conductance map $dI/dV (x,V)$ is shown in Fig. 3a.  The bound state, indicated by the horizontal arrow, extends over  2.2 $\pm$ 0.1 nm.

From the conductance map shown in Fig. 3a, another important result is established:  Close to the V$_W$ dopant, the STM detects spatial variations of the different band onsets ($\Gamma$ peak, VBM and CBM) of the 2D-TMD. It is possible to determine such band bending quantitatively, outside the 2.2-nm wide perturbation-region centered on the dopant \cite{Ref SM}. We show in Fig. 3b two spectra respectively measured at 2.50 nm (green curve) and 3.75 nm (red curve) away from the dopant. The spectra look very similar, but a shift of  30 mV is necessary to align them at best. Obviously, there is an almost rigid shift of the band structure of WSe$_2$ by +30 meV when approaching the defect by 1.25 nm. We present in Fig. 3c a plot of the shift versus the tip position x with respect to the dopant. We find an upward local band bending of the TMD (by $\sim$ 50 meV over 2 nm), demonstrating that the the V$_W$ dopant is negatively charged.

 \begin{figure}[!h]
\begin{center}
\includegraphics[width=8.5 cm,clip]{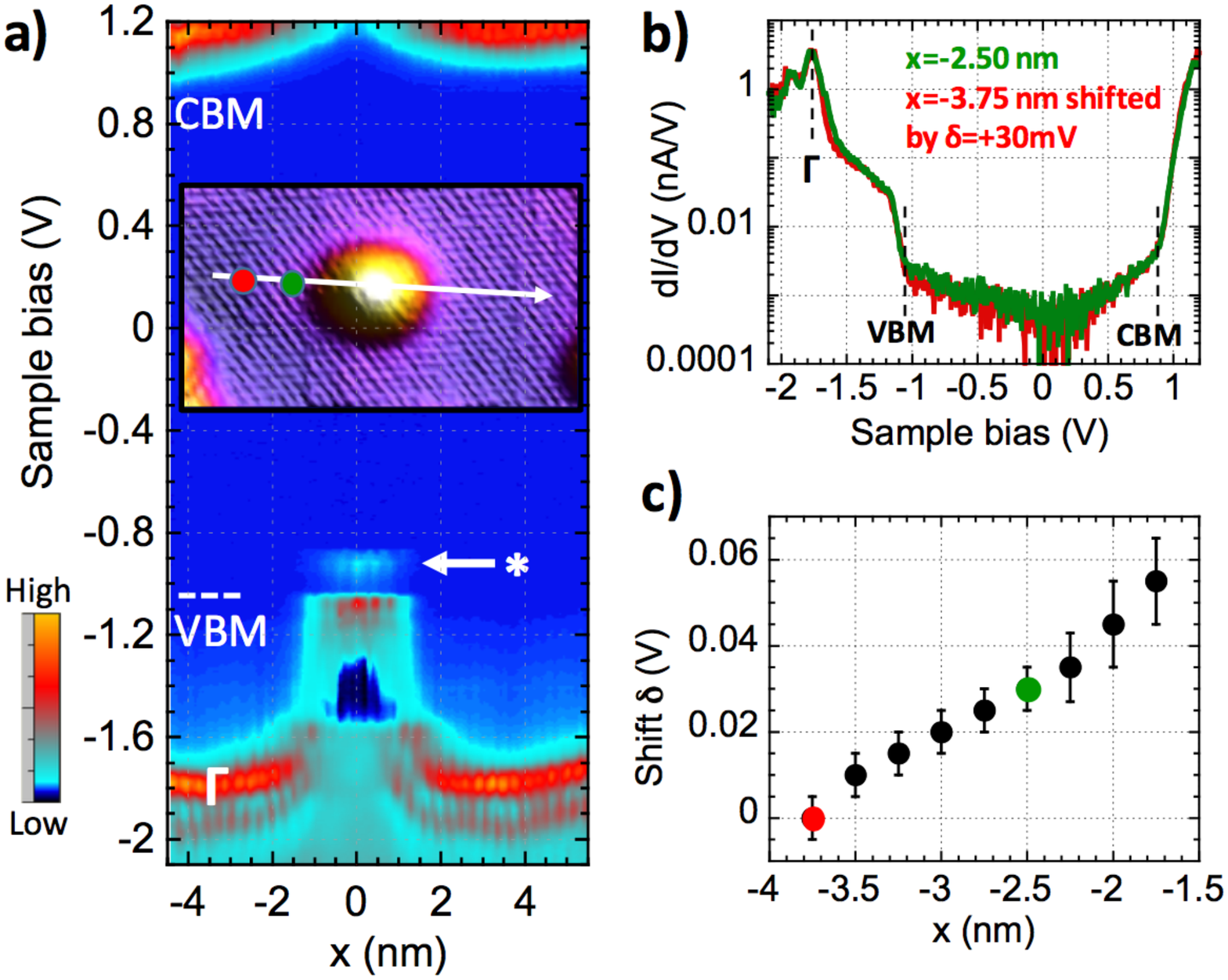}
\caption{\small STS results on a single V$_W$ dopant on 1L-WSe$_2$ on bilayer graphene.  (a) $dI/dV(x,V)$ map taken along the 10nm-long line crossing  the dopant. The energy positions of the VBM, CBM and $\Gamma$  peak are indicated on the map. The bound state  is indicated by the white arrow.  The inset shows an STM map of the area, the white line indicating the path of the tip for the map shown on (a). (b) Two $dI/dV(V)$ spectra performed at two different $x$ positions measured from the defect center, as indicated in the inset of (a). The red curve is shifted by $\delta$ = +30 mV with respect to the green curve (see text). (c) Plot of the shift $\delta$ as a function of $x$, evidencing a band bending by 50 meV over 2 nm. }
\end{center}
\end{figure}

Indeed, as reported for various surfaces of semiconductors \cite{Stroscio PRL87, Yakunin PRL04, Qui NanoLett17}, an acceptor atom is likely to trap a net negative charge, with dramatic consequences on the local electrostatic environment. In particular, the additional negative charge gives rise to an upward local bending of the SC bands, and STM images of the impurity show a smooth and isotropic apparent-height change extending over few nm, with a strongly bias-dependent contrast \cite{Stroscio PRL87, Yakunin PRL04, Qui NanoLett17}. Furthermore, the net negative charge localized at the dopant atom interacts with the holes of the valence band, in a similar way to the hydrogen-like model, leading to the formation of resonances and/or bound states, i.e. states respectively lying just below or above the VBM of the SC \cite{Ashcroft 76, Qui NanoLett17, Aghajanian SciRep18}. In Ref.  \cite{Aghajanian SciRep18}, the authors predict bound states centered on a negatively-charged defect with a radius in the nm range, together with a band bending region extending over several nm around the defect. 

Our STM/STS findings on the individual $V_W$ dopants, $\textit{i.e.}$ a band-bending region and a bound state above the VBM of the 2D-TMD, also reported recently for defects of uncontrolled origin  \cite{Schuler ACSNano19, LeQuang 2DMater18}, are fully in line with this picture of a negatively charged acceptor state. Compared to W atom, V atom is lacking one d electron, and should hence behave as an acceptor when substituting with W in WSe$_2$.  Since EG is strongly n-doped, the substrate sets the Fermi level at least 1 eV above the VBM of bare 1L-WSe$_2$ \cite{LeQuang 2DMater17}. Therefore the V-induced acceptor state is negatively charged by electronic transfer from the substrate.

To go beyond these qualitative arguments, comparison with first principle calculations is  needed. Several spin-polarized calculations have been performed for V$_{Mo}$ dopants in the cousin system 2D-MoS$_2$ \cite{Fan NanoResLett16, GonzalezDappe Nanotech16, Wu PhysLettA18, Robertson ACSNano16, Yun Condmat18, SinghUdo AdvMat17, Miao ApplSurfSci18, Mekonnen IJModPhysB18}. A common result in these works is a spin-polarized bound state above the VBM, reflecting the ferromagnetism predicted for the V-doped 2D-TMD. Similar findings were reported very recently for V-doped WSe$_2$  \cite{Duong APL19}. 
However, in most calculations the charge state of the V atom is taken neutral, which is not the case for our configuration. At variance with the neutral case,  calculation of the system with V dopants in the -1 charge state leads to zero spin polarization, suppressing magnetism in the TMD layer \cite{Robertson ACSNano16}. In Ref. \cite{Robertson ACSNano16}, the computed DOS  exhibits a resonance at the VBM and a (doubly) occupied state $\sim $ 0.15 eV above the VBM, which is in line with our results.

 \begin{figure}[!h]
\begin{center}
\includegraphics[width=8.5 cm,clip]{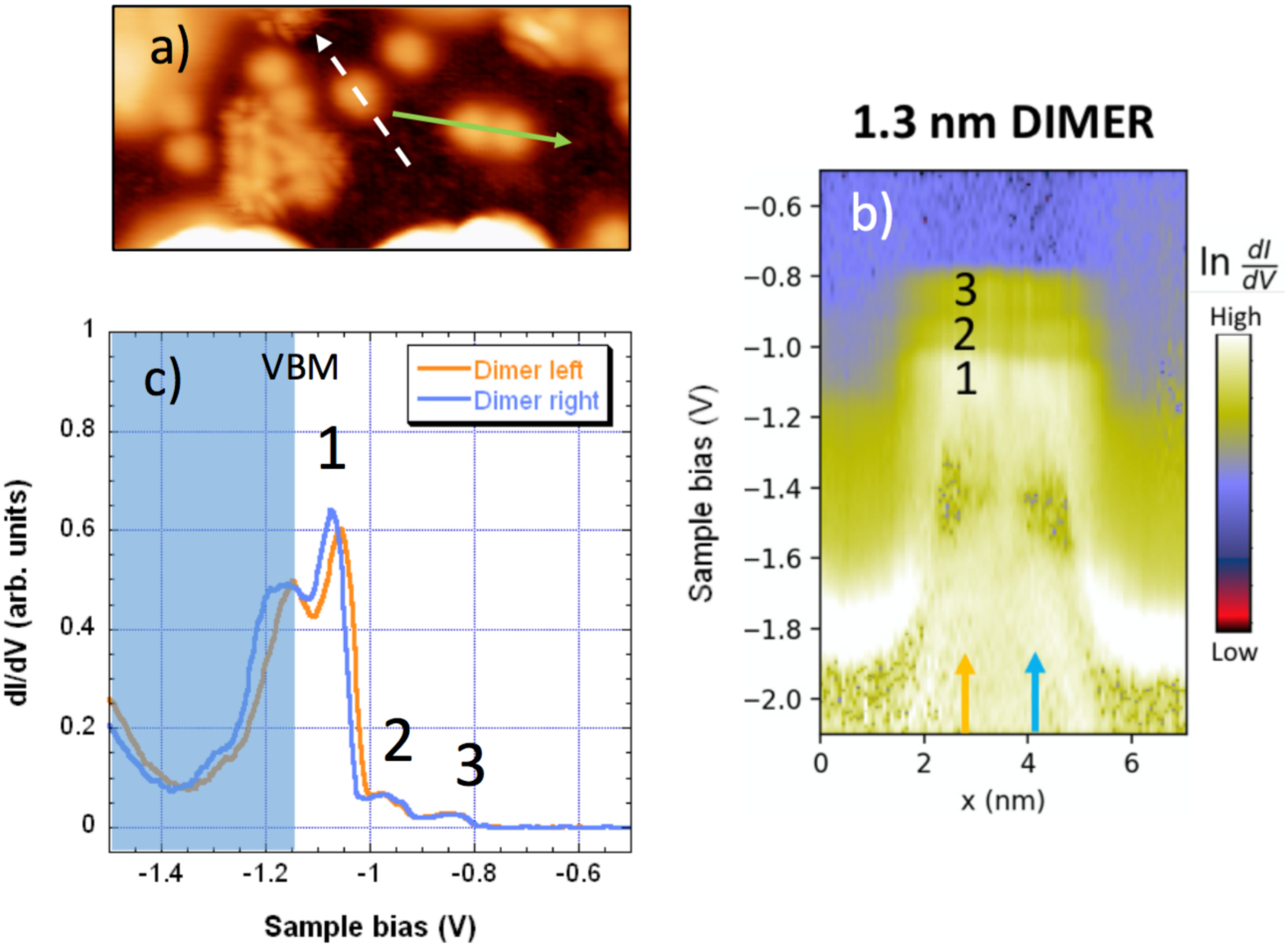}
\caption{\small Figure 4 : STS data taken on a pair of V$_W$ dopants separated by 1.3 nm. (a) Constant current STM image of the region. The 1L-WSe$_2$ flake lies on SLG.  Image size: 20.0 $\times$ 9.5 nm$^2$, $V_S$  = -1.5V. (b) $dI/dV(x,V)$ map, plotted in logarithmic color-scale, taken along the 7.0 nm-long line indicated by the green arrow on (a). Three conductance peaks are highlighted, labelled 1,2 and 3. (c) Individual spectra measured on the two lobes of the dimer indicated by the vertical arrows on (b).}
\end{center}
\end{figure}

Owing to their large apparent size, we expect a significant lateral interaction between neighboring individual V$_W$ dopants. To clarify this point, we used the STM to probe pairs of such dopants separated by short distances. As shown in Fig. 4, we collected STS data on a dimer made of two individual dopants separated by only 1.3 nm (labelled 1.3 nm dimer in the following). Such a dimer is shown in Fig. 4a, together with one individual V$_W$ dopant (located below the white arrow). We have checked in Ref.  \cite{Ref SM} that this individual dopant presents the same spectroscopic features as those reported in Figs. 2 and 3. We have then performed STS measurements along  the long axis of the 1.3 nm dimer (i.e. along the green arrow in Fig. 4a). The resulting $dI/dV(x,V)$ map is shown in Fig. 4b. The spectra measured above the 1.3 nm dimer (Figs. 4b and 4c) are dramatically different from the spectra measured above isolated V$_W$ dopants: Close to the VBM, we find three in-gap peaks (labelled 1,2,3), the highest energy peak (3) lying $\sim $ 0.3 eV above the VBM. We show in Ref. \cite{Ref SM} a very similar result for another 1.3 nm-long dimer.

These results demonstrate that the lateral interaction between two individual charged V$_W$ dopants is sizable when they are separated by 1.3 nm. However, we note that DFT calculations performed in a 4$\times$4 geometry (supercell size 1.32 nm) for a neutral V$_{Mo}$ dopant in MoS$_2$ \cite{SinghUdo AdvMat17, Miao ApplSurfSci18, Mekonnen IJModPhysB18} point to a vanishing lateral interaction, since the in-gap band above the VBM is almost flat. We believe that the long range interaction, which shows up in Figs. 4d and 4e, occurs due to the fact that the V$_W$ dopants are negatively charged in our sample.  Simple electrostatic considerations  \cite{Aghajanian SciRep18} qualitatively explain the results that we obtain for the 1.3 nm dimer. In such a configuration, the electrostatic potentials of the closely spaced dopants overlap. Hence the potential well for the holes is wider and deeper  for the dimer than for the individual dopant. As a consequence, additional and more strongly bound  hydrogenic-like in-gap states can show up \cite{Aghajanian SciRep18}, and we ascribe the peaks labelled 1,2,3 in the spectra of Fig. 4 to such electrostatically-induced in-gap states. The observation of even more strongly bound states in larger clusters of V$_W$ dopants (see Ref. \cite{Ref SM}) clearly supports the electrostatic origin of the V-induced bound states in our system.

In summary, we report here a thorough experimental study of the atomic and electronic structure of individual V dopants incorporated in monolayer WSe$_2$. Based on STM/STS techniques complemented by ADF-STEM measurements, our work shows that the V atoms are substitutional dopants at W sites of the 2D-TMD, and that they are negatively charged by a charge transfer from the n-doped graphene substrate. This net localized charge gives rise to a bound state  with a diameter of around 2.2 nm, located $\sim$140 meV above the valence band maximum of WSe$_2$. Following recent DFT calculation  \cite{Mishra  PRB13, Robertson ACSNano16, Lin JAP14}, this negative charge state of the V$_W$ dopant should impact the magnetic properties of the system, and possibly hinder any ferromagnetism in the 2D TMD layer.
An appealing perspective be would be to manipulate the charge state of the dopants by using a backgate, in order to tune the magnetism of the 2D crystal.

\section{Acknowledgements}

We thank F. Bonell, A. Marty, E. Velez and C. Vergnaud for their help during the MBE growth of the TMD samples, and L. Magaud for fruitful discussions. We acknowledge financial support by the J2D (ANR-15-CE24-0017) and the MagicValley (ANR-18-CE24-0007) projects of Agence Nationale de la Recherche.

\end{document}